\documentclass[12 pt]{report}
\usepackage{amsmath}
\usepackage{graphicx}
\usepackage{url}
\usepackage{fullpage}
\usepackage{algorithm}
\usepackage{authblk}
\usepackage{algpseudocode}
\usepackage{physics}
\usepackage[round, sort&compress, numbers]{natbib}
\graphicspath{{figures/}{ProposalFigs/}}

\title{Cytoskeletal filament length controlled dynamic sequestering of intracellular cargo}

\author{Bryan Maelfeyt}
\author{Ajay Gopinathan\footnote{agopinathan@ucmerced.edu}}

\affil{Department of Physics, University of California Merced, Merced CA, USA}

\begin{document}

\maketitle

\begin{abstract}
	The spatial localization or sequestering of motile cargo and their dispersal within cells is an important process in a number of physiological contexts. The morphology of the cytoskeletal network, along which active, motor-driven intracellular transport takes place, plays a critical role in regulating such transport phases. Here, we use a computational model to address the existence and sensitivity of dynamic sequestering and how it depends on the parameters governing the cytoskeletal network geometry, with a focus on filament lengths and polarization away or toward the periphery. Our model of intracellular transport solves for the time evolution of a probability distribution of cargo that is transported by passive diffusion in the bulk cytoplasm and driven by motors on explicitly rendered, polar cytoskeletal filaments with random orientations. We show that depending on the lengths and polarizations of filaments in the network, dynamic sequestering regions can form in different regions of the cell.  Furthermore, we find that, for certain parameters, the residence time of cargo is non-monotonic with increasing filament length, indicating an optimal regime for dynamic sequestration that is potentially tunable via filament length. Our results are consistent with {\it in vivo} observations and suggest that the ability to tunably control cargo sequestration via cytoskeletal network regulation could provide a general mechanism to regulate intracellular transport phases. 
\end{abstract}

\subsubsection{Introduction}

In eukaryotic cells, motor-driven intracellular transport is an important process that facilitates the delivery of a variety of cellular materials including cargo-loaded vesicles and organelles to different parts of the cell \cite{howardbook, ross1}. Transport occurs over relatively long distances by a combination of active motor-driven transport along protein filaments and passive thermal diffusion in the bulk \cite{arcizet1,bressloff1}. 

ATP-powered molecular motors such as kinesin, dynein, and myosin \cite{wells1, langford1, wang1} drive the active phase of transport, and carry their cargos by ``walking" along a network consisting of F-actin and microtubules, the filaments that make up the cell's cytoskeleton \cite{ross1}. The filaments are polar with distinct (+/-) ends and different motor types travel in different directions along these filaments. Myosin usually travels in the (+) direction on F-actin filaments \cite{wells1}, while kinesin and dynein both move along microtubules, with kinesin moving in the (+) direction \cite{langford1} and dynein moving towards the (-) end \cite{wang1}. When a cargo-motor complex detaches from a filament due to the unbinding of its motors, it undergoes passive diffusion in the cytoplasmic bulk phase, till it attaches to a filament again. Each cargo-motor complex, thus repeatedly attaches to, walks along and detaches from filaments until they reach their target destination \cite{bressloff1}. 

The process of cargo transport has been studied extensively from multiple perspectives including that of the molecular motors and their co-ordination \cite{malik1, klumpp1, klumpp3, beeg1, mjimuller, ando2, huang1} and features of the individual filaments they walk along \cite{jhelenius, liang1, itelley, mgramlich, lconway}. More recently, there has been a lot of work, especially theoretical, focusing on the larger scales aspects of transport including the coupling of active and passive motion \cite{ineri, pktrong, benichou, tabei1, scholz1, snider1}, the role of geometric confinement \cite{mdas,koslover} and the geometry of the cytoskeletal network itself \cite{ando1, kahana1, lconway, hafner1, hafner2}. From this last perspective, filament length, number, placement and orientation are all properties that can be regulated by a number of processes including polymerization/depolymerization, severing and nucleation at the cellular level \cite{gopinathan1,carlsson1} and have been shown to greatly affect transport first-passage times. For example, localizing the filament mass can optimize search and exit times \cite {hafner1, hafner2, ando1}, trapping regions can arise in random networks and greatly increase exit times, and orienting a small fraction of the filaments inward, towards the center of the cell can dramatically increase the mean first-passage time (MFPT) for cargo exit \cite{ando1,abel1}. If trapping regions occur naturally in random network geometries, it raises the question as to how much control the cell must exert over the geometry to avoid traps or even tunably create them if cargo sequestering is desirable. Sequestering cargo by maintaining them in motile populations that can be readily released for secretion is, for example, a defining characteristic of insulin producing pancreatic $\beta$ cells \cite{zwang} and is regulated, in part, by tuning the cytoskeletal network \cite{atomas, makalwat, zhu, bryan1}. The sensitivity of the switching between sequestering and secretion phases on the regulatory parameters is thus of critical importance in the physiological context of insulin secretion in response to glucose stimulation. Another context where the spatial localization or dispersion of cargo is important is in the color camouflage mechanism of melanophore cells that contain pigment granules carried by myosin motors along the actin cytoskeleton \cite{snider1}. These cells are able to switch between phases of aggregation near the center and dispersal throughout the cytoplasm of the pigment granules with accompanying changes in cytoskeletal geometry between the phases. Interestingly, while it was shown that switching behavior of cargos at intersections could influence the transition between the phases, the effects of the cytoskeletal changes were not really explored \cite{snider1}.  Clearly, the ability to tunably control cargo sequestration via cytoskeletal network regulation could provide a general mechanism to regulate intracellular transport phases that could apply in many different functional contexts. In this paper, using a numerical simulation approach, we explore the general question of how the existence and sensitivity of trapping regions that promote dynamic sequestering depend on the interplay of parameters governing the cytoskeletal network geometry, in particular filament lengths and polarization (inwards/outwards).\par

% In our work, for simplicity, we will assume one motor type, and thus, one type of filament and polarization. A filament's orientation (or ``polarization") is determined by its (+) end and (-) end.
%Points: a. phase transition of survival probability, large fluctuations - note these are network=network standard deviations. b. Sensitivity to survival prob as fn of length at fixed polarization. For eg depolymerization can drive the dissolution of traps and hence export. 
% hybrid formalism combines explicit filaments and probability density description of cargo. Coarse grained models of filaments have a hige draw-back; memory (refer prev paper). Taking into account explicit filaments becomes important. Single cargo simulations on such networks are fast but capturing FPTD requires  a large number and the noise increases at large times where events are rare and statistics are poor - this is avoided by PDE which allows us to compute FPTD accurately upto any desired time with limitations due to munerical accuracy.
%optimal cargo sequestration; dynamic pools of cargo
%  what about with no exit? enhance  these pools - non secretory cargo - like melanophores
% polarization at transition depends on what time you measure survival probability, need to show at different times? will be higher at  smaller times
%

There has been much computational work done through the use of simulations and numerical analysis in order to understand the intracellular transport process better. There are two broad classes of computational approach - (i) explicitly simulating the dynamics of a single cargo and (ii) time evolution of differential equations describing the spatial distribution of an ensemble of cargo. Explicit simulations of cargo movement typically rely on a coarse grained description of filament effects. One type of simulation, for example, involves the use of random velocity models \cite{ahafner} to account for ballistic transport along filaments and use this to model the spatial inhomogeneity of physical cytoskeletal networks \cite{kschwarz}. Still, other methods focus on drawing cargo binding rates and movement information from distribution functions \cite{kchen}. However, the presence of explicit filaments in models makes a qualitative difference allowing for the possibility of trapping regions, memory effects due to filament rebinding and significant changes in mean transport times \cite{ando1, bryan1}. 

A different approach is to consider the evolution of the probability distribution of a cargo ensemble. Systems of differential equations can model the time evolution of cargo spatial distributions. These tend to require the coupling of both the passive diffusion and active transport \cite{ineri} phases. A particularly simple and interesting limit of this problem occurs when filaments are aligned and motor-diven transport in the active phase facilitates advective transport in the ``passive" phase \cite{pktrong}. In such methods, as compared to simulations of individual cargo dynamics, there is a trade-off of not requiring extensive sampling for noise reduction at the cost of precision in numerical integration upto late times. Such an approach allows for the accurate evaluation of mean first passage times but on the other hand cannot be used to evaluate stochastic variations in cargo first passage times. As is the case with most cargo dynamics simulation methods, these models typically do not use explicit filaments in their calculations, which produces qualitative differences as pointed out above.\par

Here we combine the probability distribution approach with an explicitly represented inhomogeneous cytoskeletal network whose filaments are randomly oriented in two dimensions. This allows us to capture both the active and passive of transport through numerical integration by treating individual cargos as random walkers in the continuum limit \cite{avraham1} and incorporating switching between the active and passive phases by solving coupled differential equations \cite{bressloff1} that describe the time evolution of spatial cargo distribution both on and off the explicit cytoskeletal networks. These {\it in silico} networks are generated by placing filaments, represented by straight lines with a defined polarity (given by the direction of active ballistic transport), at random locations and orientations within the cytoplasm. Very recently, a similar approach \cite{holmes}, but with apolar filaments where the active and passive transport are both treated as subdiffusive, was used to show that altering the microtubule network could regulate secretion of insulin granules by withdrawing them from the periphery and inhibiting their ability to fuse with the cell membrane.  Previous results \cite{ando1, abel1}, using explicit simulations of single cargo with ballistic transport on polar filaments, have shown that even with fixed orientations and locations, simply having a modest fraction of filaments polarized towards the interior of the cell, produces trapping regions and greatly slows down cargo transport. Given its potential functional relevance, we are particularly interested in whether such trapping regions can function as dynamic and tunable regions for cargo sequestration. We therefore systematically examine the dependence of the survival probability of cargo (probability that the cargo has not reached the outer boundary) at a fixed time, on both filament length and polarization. In order to isolate the effects of these parameters, we choose the simplest model with these features and neglect other effects present {\it in vivo} including viscoelastic interactions \cite{mdawson}, filament elasticity \cite{acaspi,holmes}, multiple motors and switching \cite{snider1} and confinement \cite{koslover, mdas}. We find that, as expected, increasing the polarization of the filaments towards the interior results in increased survival probability. However, we find that the survival probability is non-monotonic with increasing filament length, indicating an optimal regime for dynamic sequestration that is potentially tunable via filament length. In this paper, we examine the origins of this behavior and assess its tunability.

\subsubsection{Methods}

Our simulation domain consists of a circular cell with an outer radius of 10 $\mu$m and an inner nuclear boundary with a radius of 5 $\mu$m \cite{ando1, bryan1}. We describe the transport process by considering the time evolution of $P(x,y,t)$, the 2D probability distribution function of cargos as a function of position and time.  Within the annular cytoplasm, we also place a randomized network of explicit cytoskeletal filaments, which are straight lines of fixed length with random locations and orientations (in continuous space, see \cite{ando1,bryan1} and SI for more details on network generation). Filament center of masses are shifted radially to ensure that they lie completely within the domain.  The physical processes we wish to describe involve the dynamics of cargos carried by kinesin motors moving ballistically while on filaments with a speed of $v = 1 \mu m/s$, and subject to diffusion while off filaments with a diffusion constant of $D = 0.051 \mu m^2 /s$.  Cargos should also be able to bind and unbind filaments at rates of $k_{on} = 5 s^{-1}$ and $k_{off} = 1 s^{-1}$ (set by kinesin binding and unbinding rates \cite{bryan1}), respectively. 

%Bryan Maelfeyt added this below (Feb. 11 2019)
% Now, the discrete space that the cell occupies is a 2D lattice with lattice site locations $(x,y)$ $0.1 \mu m$ apart. The cargo distribution can exist at any $(x,y)$ location within the cell. The filament endpoints are now placed randomly throughout the cell and the filament itself is made to be $0.2 \mu m$ thick to account for the cargo radius, within which, the distribution has a chance of attaching to a filament.

During the diffusive transport phase, we model individual cargos as random walkers. For a distribution of cargos, we can then describe its time evolution by \cite{avraham1}

\begin{equation}
\frac{\partial P(x,y,t)}{\partial t} = D \nabla^2 P(x,y,t).
\label{eq:diffusion}
\end{equation}

where $P(x,y,t)$ is the probability distribution function of cargos as a function of position and time and $D$ is the diffusion constant. Because the cargo must move via diffusion off filaments, and ballistic motion while on filaments, we can model the transport dynamics as a combination of diffusion and constant drift \cite{bressloff1},  

\begin{equation}
\frac{\partial{P}}{\partial{t}} = - (\nabla \cdot \vec{v}) P + D \nabla^2 P,
\label{eq:advectionDiffusion}
\end{equation}

where $\vec{v}$ is the velocity of cargos during active transport along the filaments, with the direction of the velocity being set by the direction of polarization of the filament, which is assumed to be fixed in time for a given filament.

Thus the ``on" and ``off" phases of motion are distinct and well-defined. Given this, we break up \eqref{eq:advectionDiffusion} into two equations, one corresponding to a distribution of cargo \emph{on} ($P_{on}$) filaments and one corresponding to an \emph{off} distribution ($P_{off}$). The active and passive phases of transport can then, respectively, be represented by

\begin{equation}
\frac{\partial{P_{\text{on}}}}{\partial{t}} = - (\nabla \cdot \vec{v}) P_{\text{on}}
\end{equation}

and

\begin{equation}
\frac{\partial{P_{\text{off}}}}{\partial{t}} = D \nabla^2 P_{\text{off}}
\end{equation}

We also have switching between these two phases of motion.  For a distribution that switches between these two states \cite{bressloff1}, we can write

\begin{align}
\frac{\partial{P_{\text{on}}}}{\partial{t}} &= - (\nabla \cdot \vec{v}) P_{\text{on}} - k_{\text{off}} P_{\text{on}} + k_{\text{on}} P_{\text{off}},\\
\frac{\partial{P_{\text{off}}}}{\partial{t}} &= D \nabla^2 P_{\text{off}} + k _{\text{off}} P_{\text{off}} - k_{\text{on}} P_{\text{off}},
\label{eq:totalProb}
\end{align}

where, at all times, the total probability distribution is given by

\begin{equation}
\frac{\partial{P}}{\partial{t}} = \frac{\partial{P_{\text{on}}}}{\partial{t}} + \frac{\partial{P_{\text{off}}}}{\partial{t}}.
\end{equation}

Here, $k_{on}$ and $k_{off}$ are the cargo attachment and detachment rates, respectively, which couple the two differential equations. In order to implement a numerical scheme to solve the differential equations, we first discretize the annular domain using a square lattice of lattice constant $0.1 \mu m$, and keeping track of $P_{on/off}(x,y,t)$ at the lattice sites. The filaments are then discretized by considering all lattice sites within $0.1 \mu m$ of the filament lines to be `filament' sites. This size reflects the spatial range within which the cargo can attach to filaments, which is set by the cargo radius, $c_{rad} = 0.1 \mu m$.
Our goal is to find a numerical solution to \eqref{eq:totalProb} at each successive point in time as we let the distribution evolve. We begin by approximating, to first-order, the differential equations as

\begin{align}
\frac{P^{n+1}_{on,i,j} - P^{n}_{on,i,j}}{\Delta t} \approx &- \frac{v_x}{2 \Delta x}(P^{n}_{on,i+1,j} - P^{n}_{on,i-1,j}) - \frac{v_y}{2 \Delta y}(P^{n}_{on,i,j+1} - P^{n}_{on,i,j-1})\nonumber \\
&+ (k_{on} P^{n}_{off,i,j} - k_{off} P^{n}_{on,i,j})
\label{eq:diffApprox}
\end{align}

and

\begin{align}
\frac{P^{n+1}_{off,i,j} - P^{n}_{off,i,j}}{\Delta t} \approx &+ \frac{D}{{\Delta x}^2}(P^{n}_{off,i+1,j} + P^{n}_{off,i-1,j} - 2 P^{n}_{off,i,j})\nonumber \\
&+ \frac{D}{{\Delta y}^2}(P^{n}_{off,i,j+1} + P^{n}_{off,i,j-1} - 2 P^{n}_{off,i,j})\nonumber \\
&- (k_{on} P^{n}_{off,i,j} - k_{off} P^{n}_{on,i,j}).
\label{eq:ballApprox}
\end{align}

Here, $P^{n}_{i,j}$ is the distribution at position $(i,j)$ in space, at time step $n$. $P^{n+1}_{i,j}$ will then be the distribution at the next time step ($n + 1$).  $\Delta x$ and $\Delta y$ are the distances between points in space and $\Delta t$ is the size of the time step. $v_x$ and $v_y$ are the velocity components representing the speed at which the cargo moves while on a filament, which is set by the motor type and direction of filament polarization.

To implement our integration, we will, at each point in space (for each time step in the integration), update the probability distribution. To do this, we first allow the distribution to either ``attach" or ``detach from the network. This gives us updated values for $P_{off}$ and $P_{on}$:

\begin{align}
P^{n+1}_{off,i,j} &= P^{n}_{off,i,j} + \Delta t \cdot(-k_{on} P^{n}_{off,i,j} + k_{off} P^{n}_{on,i,j}),\nonumber \\
P^{n+1}_{on,i,j} &= P^{n}_{on,i,j} + \Delta t \cdot (k_{on} P^{n}_{off,i,j} - k_{off} P^{n}_{on,i,j}).
\label{eq:switch}
\end{align}

We then implement movement both off and on the filaments using,

\begin{align}
P^{n+1}_{off,i,j} = &P^{n+1}_{off,i,j} + \Delta t \cdot (\nonumber \\
&+ \frac{D}{{\Delta x}^2}(P^{n}_{off,i+1,j} + P^{n}_{off,i-1,j} - 2 P^{n}_{off,i,j})\nonumber \\
&+ \frac{D}{{\Delta y}^2}(P^{n}_{off,i,j+1} + P^{n}_{off,i,j-1} - 2 P^{n}_{off,i,j})),
\label{eq:moveOff}
\end{align}

 and

\begin{align}
P^{n+1}_{on,i,j} = &P^{n+1}_{on,i,j} + \Delta t \cdot (\nonumber \\
&- \frac{v_x}{2 \Delta x}(P^{n}_{on,i+1,j} - P^{n}_{on,i-1,j})\nonumber \\ 
&- \frac{v_y}{2 \Delta y}(P^{n}_{on,i,j+1} - P^{n}_{on,i,j-1})).
\label{eq:moveOn}
\end{align}

It is to be noted that, in \eqref{eq:moveOn}, $P_{on}$ can only be nonzero where a filament exists at $(i,j)$. We take this into account explicitly using our knowledge of the locations of the discretized filament sites and furthermore assume that the distribution ``walks" off the ends of filaments (see Appendix for details of this implementation).  

After properly updating $P_{off}$ and $P_{on}$, we can calculate the total probability distribution at each point in space for each time step $n$,

\begin{align}
&P^{n}_{off,i,j} = P^{n+1}_{off,i,j}\nonumber \\
&P^{n}_{on,i,j} = P^{n+1}_{on,i,j}\nonumber \\
&P^{n}_{i,j} = P^{n}_{off,i,j} + P^{n}_{on,i,j}.
\end{align}

 At every instance in time, we can therefore determine the probability that cargo has stayed within the cell i.e. the survival probability, $S(t)$, by integrating $P$ over its spatial domain (the interior of the cell).  The rate at which the survival probability decreases in time gives us the first-passage time distribution (FPTD) ($F(t)$ below):

\begin{equation}
S(t) = \int_{\text{domain}} P(x,y,t) dx dy,
\label{eq:S}
\end{equation}

\begin{equation}
F(t) = - \frac{\partial{S(t)}}{\partial{t}}.
\label{eq:F}
\end{equation}

Finally, the mean of the FPTD gives us the mean first passage time (MFPT),

\begin{equation}
\text{MFPT} = \int_0^{\infty} t F(t) dt.
\label{eq:mfpt}
\end{equation}

\begin{figure}[h!]
\centering
\includegraphics[scale =0.5]{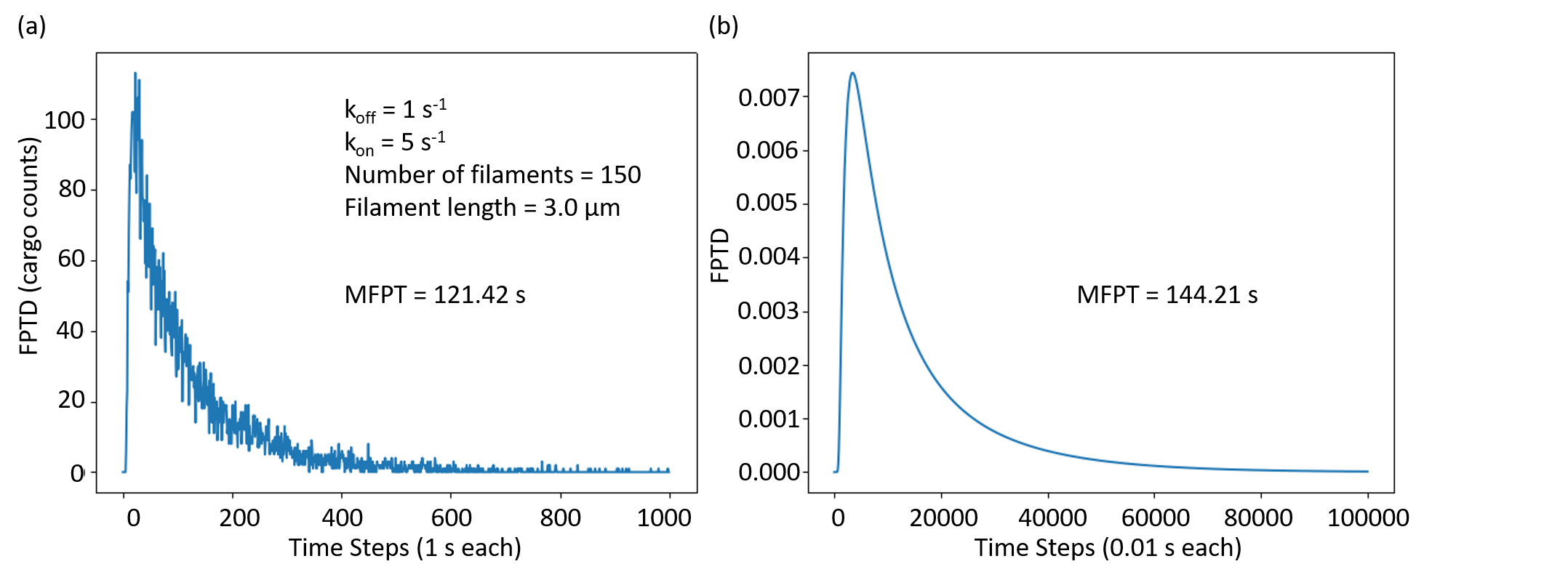}
\caption{\label{fig:fptds} A comparison of FPTD achieved via (a) simulation of 10000 cargos and (b) numerical integration. Notice that the MFPTs are comparable and that the FPTD in (b) is smoother.}
\end{figure}

In practice, we only integrate \ref{eq:mfpt} to the time where the probability distribution leaving the cell is smaller than the desired numerical accuracy. A comparison of the FPTDs and MFPTs obtained through numerical integration with those which are obtained through the simulation of the transport of multiple cargos (Fig. \ref{fig:fptds}) for similar parameter values, shows their distinctive features. For ease of comparison, they have been deliberately plotted side-by-side with the axes values reflecting the typical measurements associated with these different methods.

Fig. \ref{fig:fptds}a shows a FPTD obtained through the simulation of the movement of 10000 cargos. The counts on the y-axis are the actual numbers of cargo that pass the boundary in any given time interval. It can clearly be seen that there is some noise inherent in the simulation itself which reflects the stochastic nature of the underlying process. This is different from what can be seen in Fig. \ref{fig:fptds}b, where the FPTD shown was obtained by numerical integration as outlined above. The x-axis is now in units of discretized time steps while the y-axis shows the probability of having the corresponding first passage time.  One can see that the MFPT is similar in both cases as are the typical values of FPTD . For example, at $t=20$s, FPTD cargo counts is about 18 (out of 10000 total), while the FPTD value from the numerical method is 0.0018 at 20000 time steps (of 0.01s each). It is to be noted that the FPTD from the numerical integration is completely smooth, which highlights the speed and accuracy of the numerical integration method in the limit of infinite cargo. It is to be noted that this is at the expense of information about individual trajectories and intrinsic variability of passage times for small cargo numbers that can be obtained from the explicit simulation method.

\subsubsection{Probability Distribution Evolution on Networks of Different Polarization Biases}

\begin{figure}[h!]
\centering
\includegraphics[scale =0.5]{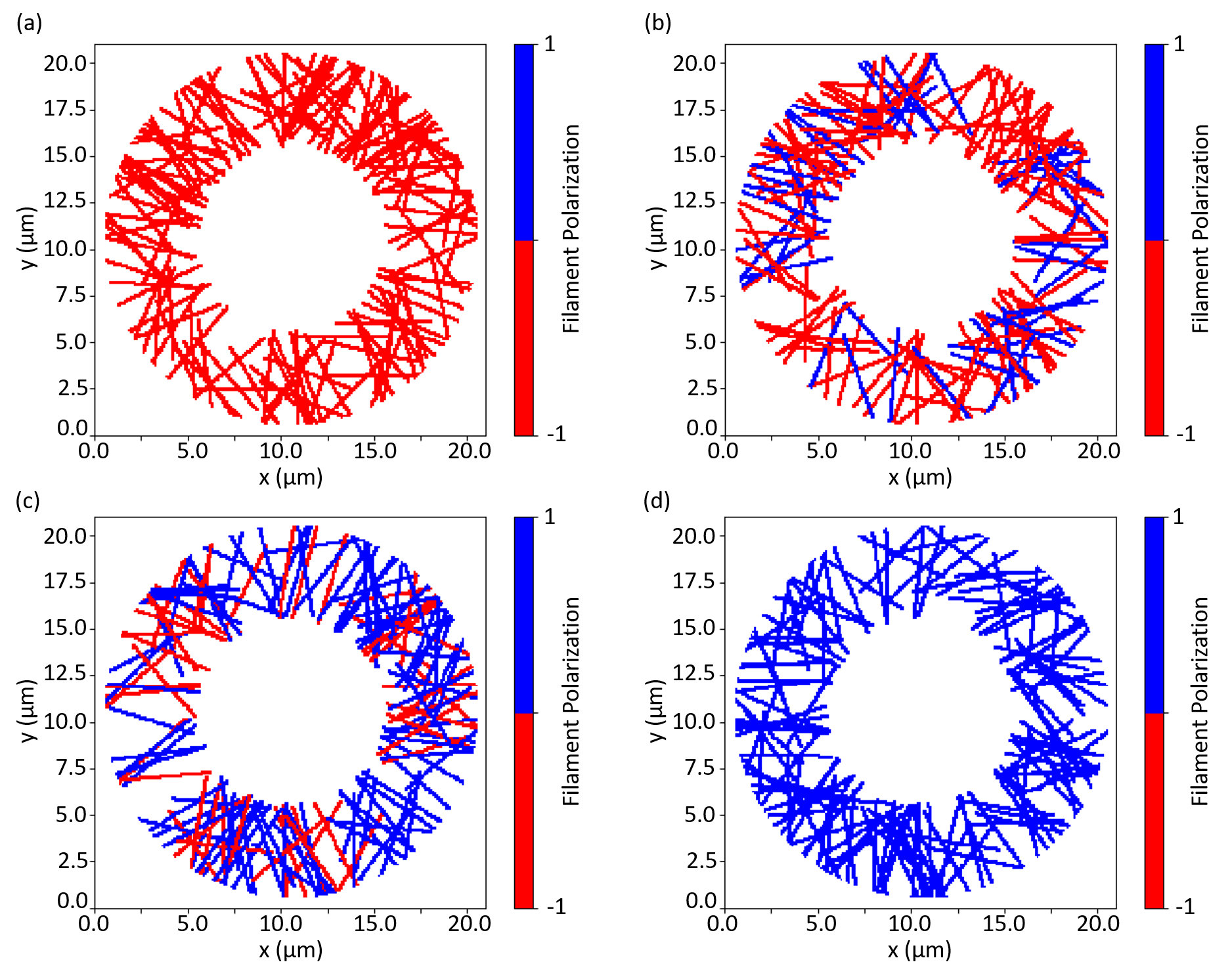}
\caption{\label{fig:pols100} Different polarization biases for 150 filaments, each with a length of 5 $\mu$m. The polarization biases are (a) 0.0 (0 \% of filaments pointing outward), (b) 0.3 (approximately 30 \% of filaments pointing outward), (c) 0.7, and (d) 1.0.}
\end{figure}

 We define the polarization of an individual filament as a binary quantity with value +1 if the filament has its plus end (end toward which motors in our context move) closer to the outer membrane than its minus end and -1 otherwise. When the network is constructed each filament has a random angle with respect to the radial direction and can point either inward or outward.  We define the {\it network polarization bias} as the probability that each filament in the network has of being polarized outward, towards the cell membrane (away from the nucleus).Thus, a network with a polarization bias of 0.1 will have about 10\% of its filaments polarized toward the cell membrane (i.e., having a polarization of +1), and the remainder ($\sim $ 90\%) of its filaments polarized towards the nucleus (having a polarization of -1). Fig. \ref{fig:pols100} shows four different networks with four different polarization biases. Each network contains 150 filaments, each with  length of 5 $\mu$m.

Filaments in the figure are colored red if they are polarized inwards (polarization = -1) and blue if they are directed outwards. In Fig. \ref{fig:pols100}a, all filaments (colored red) have a polarization of -1 and the polarization bias for the network is 0.0.  Fig. \ref{fig:pols100}b shows a network with a polarization bias of 0.1, obtained by allowing each filament, as it is generated and laid down, to have a 0.1 probability of pointing outward. The networks in Figs. \ref{fig:pols100}c and \ref{fig:pols100}d are generated similarly, with Fig. \ref{fig:pols100}c showing a network with approximately 70\% of its filaments having a polarization of +1, and Fig. \ref{fig:pols100}d showing a network with 100\% of its filaments having a polarization of +1.

\begin{figure}[h!]
\centering
\includegraphics[scale =0.5]{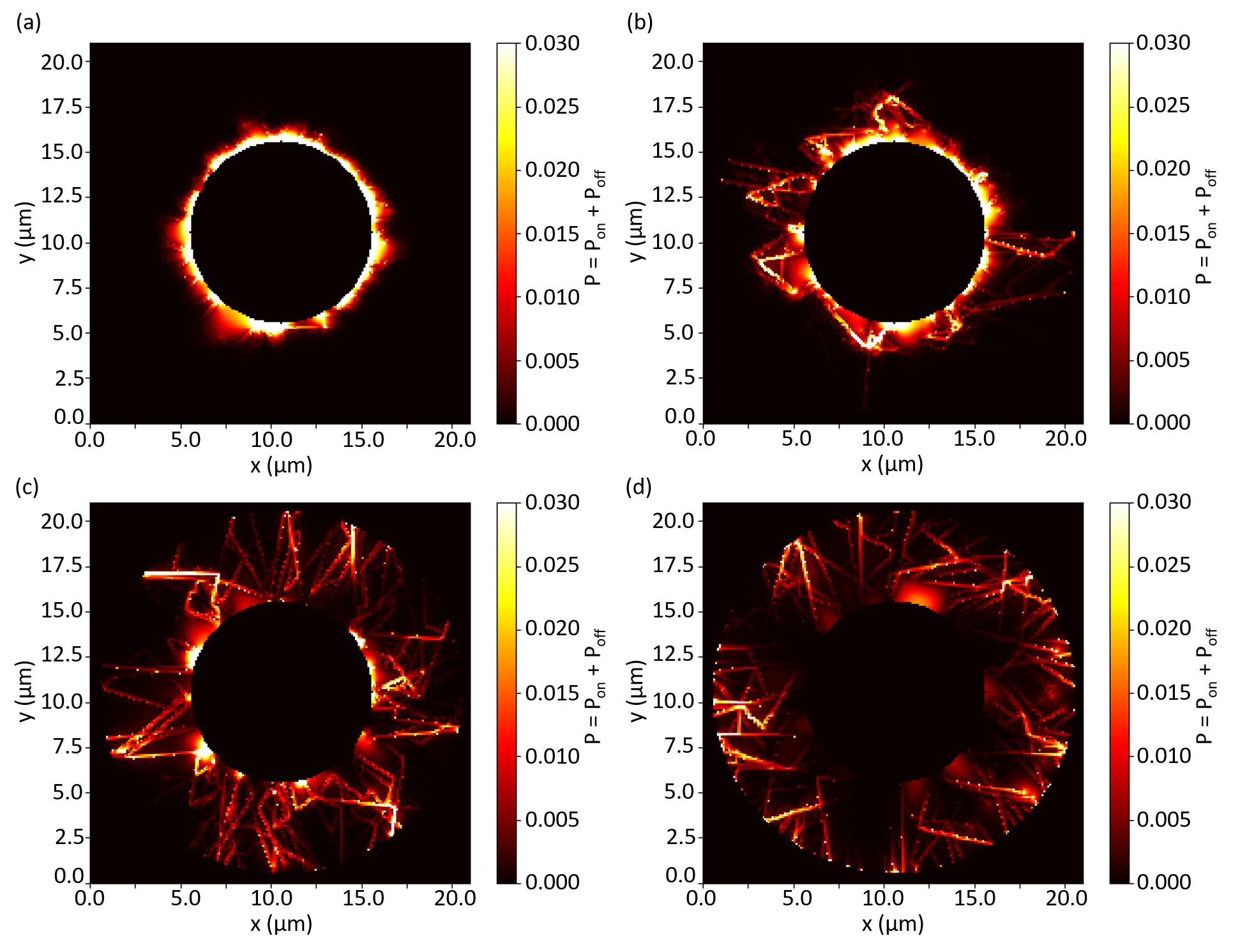}
\caption{\label{fig:probDist100} State of the cargo distribution after moving (for 100 s) on and off a cytoskeletal network comprised of 150, 5 $\mu$m filaments for polarization biases of (a) 0.0, (b) 0.3, (c) 0.7, and (d) 1.0. The distribution is evolved over the networks in Fig. \ref{fig:pols100}}
\end{figure}

To gain some insight into how networks of different polarization biases affect the time evolution of the probability distribution of cargo, we consider the  distribution at intermediate times that are comparable to the time required to traverse the cell via pure diffusion($\sim 100$ s). In keeping with how the positions of explicit individual  cargos are initialized in \cite{ando1}, here, the cargo distribution begins as an annulus of width 0.2 $\mu$m near the surface of the nucleus, in the off state. We then allow the distribution to evolve in time according to equations \ref{eq:switch}, \ref{eq:moveOff} and \ref{eq:moveOn}. Fig \ref{fig:probDist100} shows the state of the distribution after 100s on the four different networks shown in Fig. \ref{fig:pols100}.

As one would expect, an increasing polarization bias enhances the fraction of the cargo probability distribution that reaches the cell membrane and exits. For example, in Fig. \ref{fig:probDist100}a, where the polarization bias is 0.0, much of the distribution is still near the nucleus, which indicates that the MFPT for cargo distribution on this network will be very high, while in Fig. \ref{fig:probDist100}d, where the bias is 1.0, the distribution is located more toward the outer membrane and is overall lower in magnitude, signifying a reduction in MFPT. Thus increasing filament polarization with fixed length filaments clearly accelerates export and should result in a monotonic reduction in MFPT.

\subsubsection{Survival Probability as a Function of Filament Length and Polarization}

\begin{figure}[h!]
\centering
\includegraphics[scale =0.5]{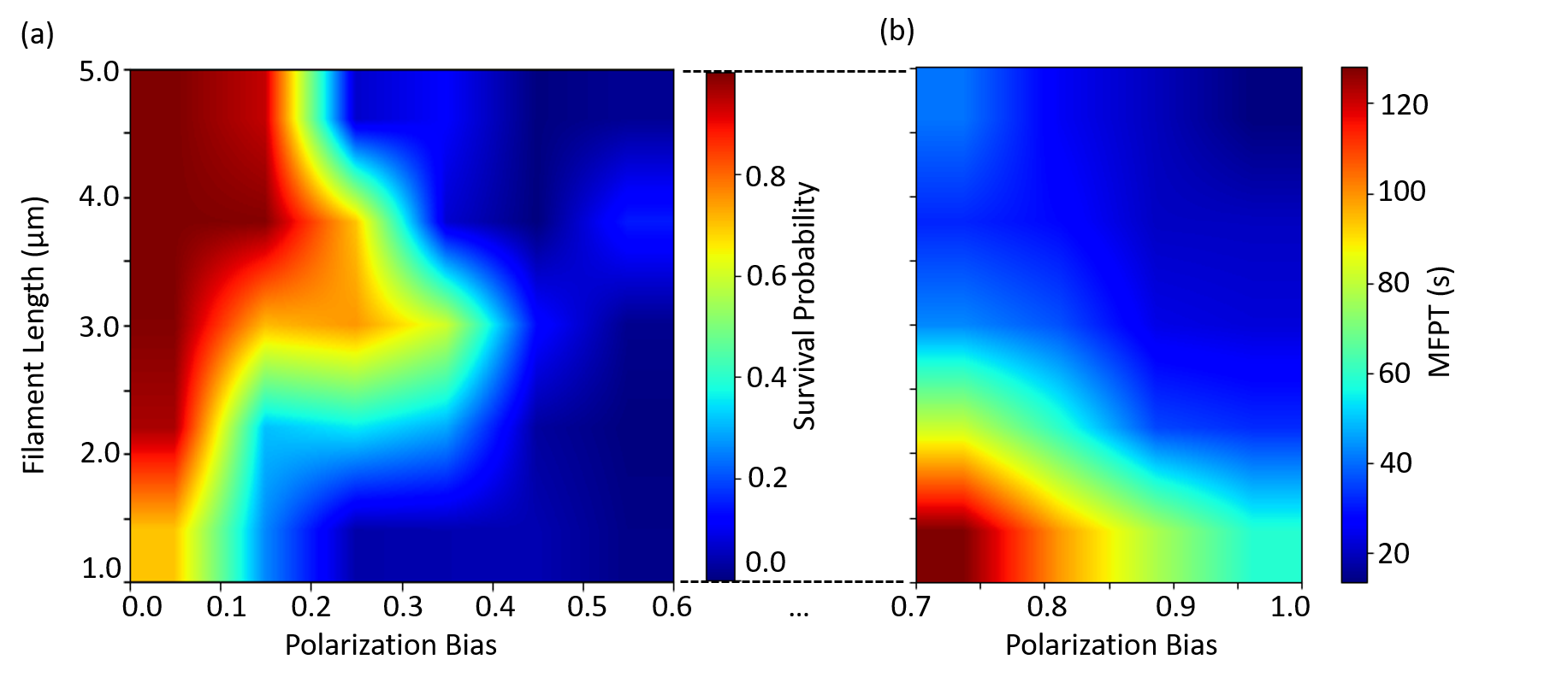}
\caption{\label{fig:mfpts} (a) Survival probability at 1000 s as a function of filament length and polarization bias for networks with 150 filaments. Above a polarization bias of approximately 0.7, the survival probability is negligibly small. (b) The MFPT for different filament lengths and polarization biases in the regime where the survival probability is close to zero.}
\end{figure}

In order to quantify the efficiency of sequestration/export as a function of network geometry, we focus on how the MFPT and survival probability at late times are affected by filament length and network polarization bias. To do this, we allow the the initial probability distribution to evolve for 1000 s over networks of different combinations of filament lengths and polarization biases. This time is an order of magnitude longer than the diffusion timescale of the previous section (MFPT $\sim 140$ s), allowing us to access distributions at later times but short enough that we can study the survival probability and visualize the spatial distribution of cargo to understand the interplay between sequestering in traps and export.   We can see in Fig. \ref{fig:mfpts}a that for filament lengths of 1, 2, 3, 4, and 5 $\mu$m, and for network polarization biases of 0.0, 0.1, 0.2, 0.3, 0.4, the survival probability is mostly nonzero and even approaches 1.0 for low polarization biases, allowing us to monitor changes in the survival probability. This is notable, as the time  is an order of magnitude greater than the diffusion timescale and we would expect an exponentially small probability distribution at such late times in the absence of filaments. The inward funneling of cargo by filaments at low polarization biases thus has a dramatic effect. As indicated in the previous section as well, the survival probability decays monotonically with increasing polarization bias as more and more filaments point outwards. However, there is a non-monotonic trend in the survival probability with filament length at intermediate polarizations (0.2-0.3), with a maximum occurring around $3 \mu$m, where the transition from high to low survival probaility occurs at higher polarization biases. We will return to this point in the following sections.

 It is important to note that an accurate determination of the MFPT requires a characterization of the full FPTD out to very late times (i.e., when the survival probabiilty is low and the FPTD is near zero). Beyond a polarization bias of 0.7 and for all filament lengths, the survival probability is essentially zero (within our numerical accuracy), meaning that MFPT calculations are accurate. The MFPT is therefore plotted in Fig. \ref{fig:mfpts}b for these polarization biases. The results indicate what one would expect based on our findings from the previous section. As the polarization bias for the network increases, the MFPTs decrease in value. The same is true for increasing filament length as well signifying that longer filaments lead to more efficient export at higher polarization biases. 

\subsubsection{Intermediate Filament Lengths and Polarization Biases Enhance Survival Probability}

\begin{figure}[h!]
\centering
\includegraphics[scale =0.4]{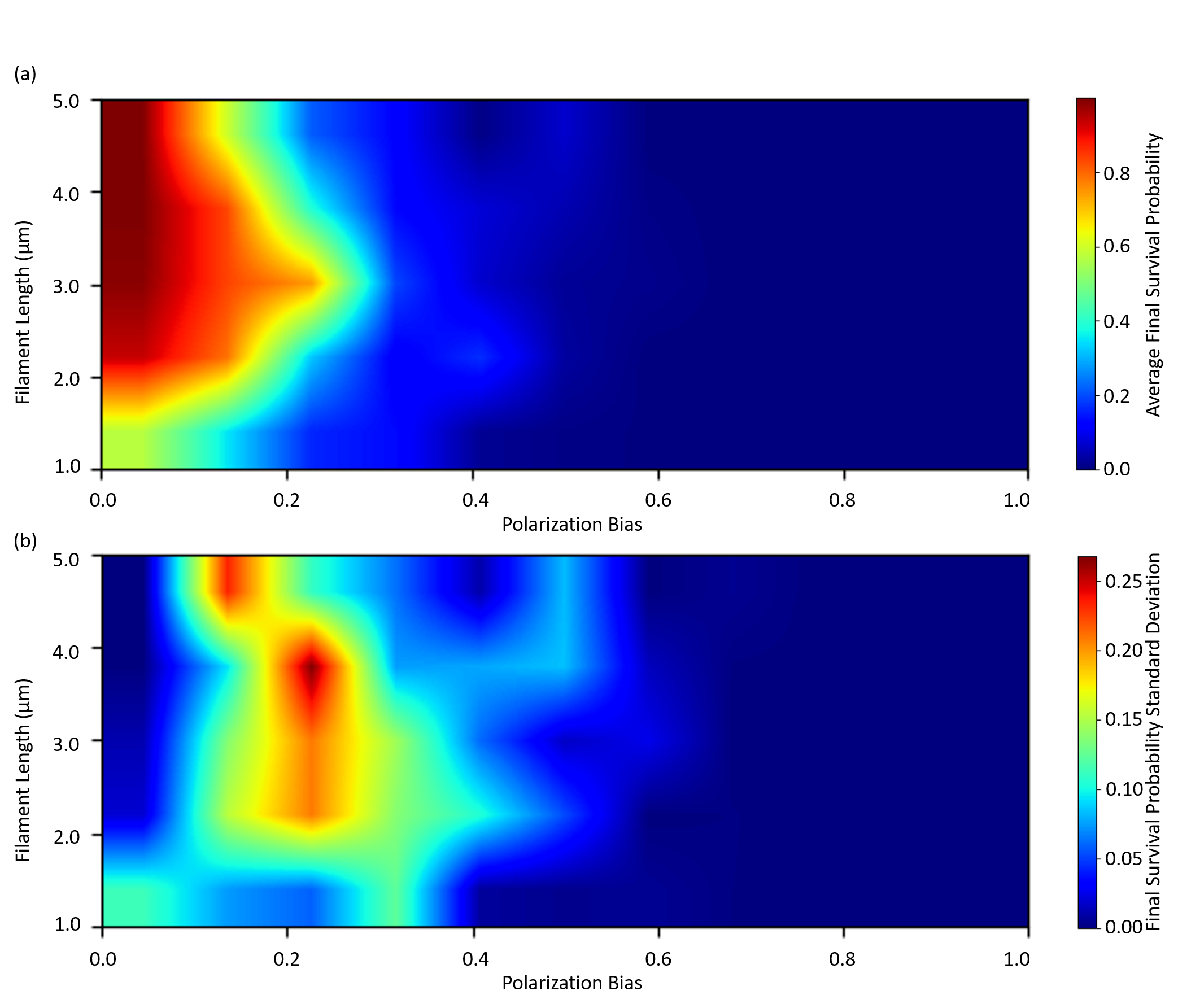}
\caption{\label{fig:survivalAvg} (a) The survival probability averaged over five different networks at each filament length, polarization bias. (b) The network to network standard deviation of the survival probability.}
\end{figure}

In order to make sure that the filament length/polarization bias effects that we referred to in the previous section are not just  artifacts of the particular network geometries simulated, we calculate the survival probability for the probability distribution at 1000 s on five different networks at each filament length/ polarization bias combination (filament lengths of 1, 2, 3, 4, and 5 $\mu$m, and polarization biases of 0.0, 0.1, 0.2, 0.3, 0.4, 0.5, 0.6, 0.7, 0.8, 0.9, and 1.0) and average the results. We can see in Fig. \ref{fig:survivalAvg}a that the transition of the survival probability from high to low values has the same general features as in \ref{fig:mfpts}a , including the non-monotonicity with filament lengths with a maximum around a filament length of 3 $\mu$m.

In order to characterize the sensitivity of the survival probability to the network parameters near this transition, we plot in Fig. \ref{fig:survivalAvg}b, the standard deviation in the survival probability across networks. We see that fluctuations are largest near the transition region indicating a higher susceptibilty to changes in filament length/polarization. To further demonstrate this point, we plot the average survival probability as a function of network polarization bias for different filament lengths. We note in particular that, at a filament length of 3 $\mu$m and a polarization bias of 0.2, the survival probability is greater than for any other filament length at this polarization bias, and that the standard deviation (given by the size of the error bars) is relatively large as well.

These results are consistent with the findings of \cite{ando1}, where, as filament polarizations were changed from +1 to -1 when the filaments were 3 $\mu$m in length, significant increases in MFPT were found. Here, we quantify this transition by monitoring the survival probability which, as we can see from \eqref{eq:S}, \eqref{eq:F}, and \eqref{eq:mfpt}, is directly related to the MFPT.

\begin{figure}[h!]
\centering
\includegraphics[scale =0.5]{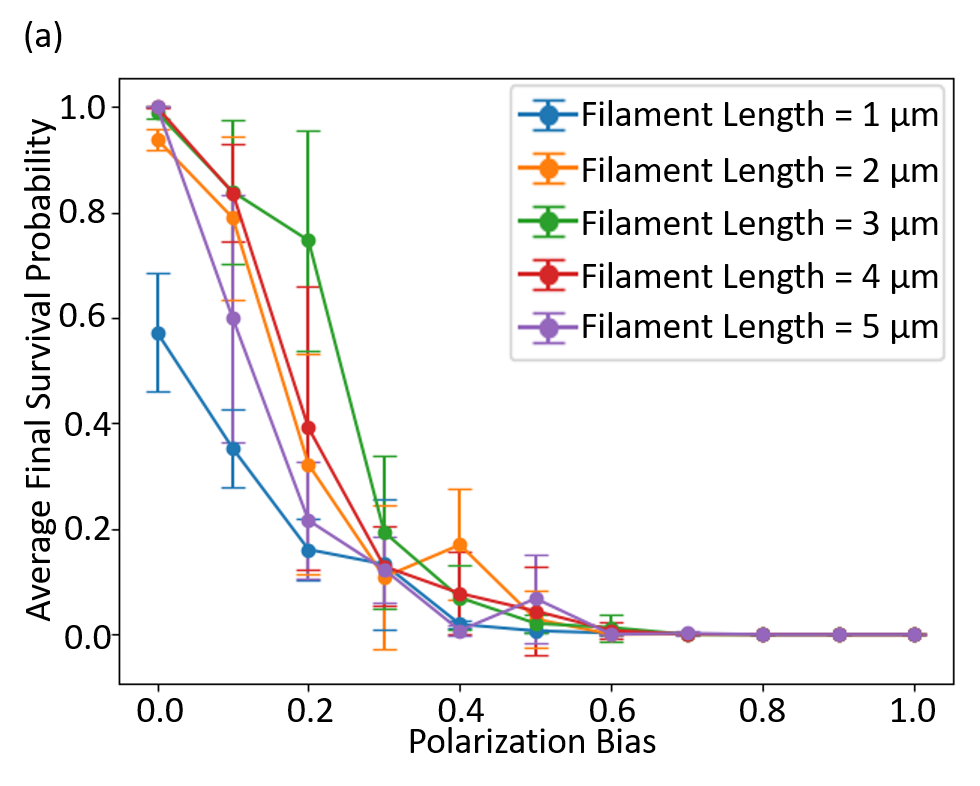}
\caption{\label{fig:survivalAvgLines}The average survival probability, as in Fig. \ref{fig:survivalAvg}a, but for different filament lengths, as a function of polarization bias. The error bars are the standard deviations calculated for Fig. \ref{fig:survivalAvg}b.}
\end{figure}

\subsubsection{Intermediate Filament Lenghts and Polarization Biases Facilitate Sequestering in the Bulk}

\begin{figure}[h!]
\centering
\includegraphics[scale =0.5]{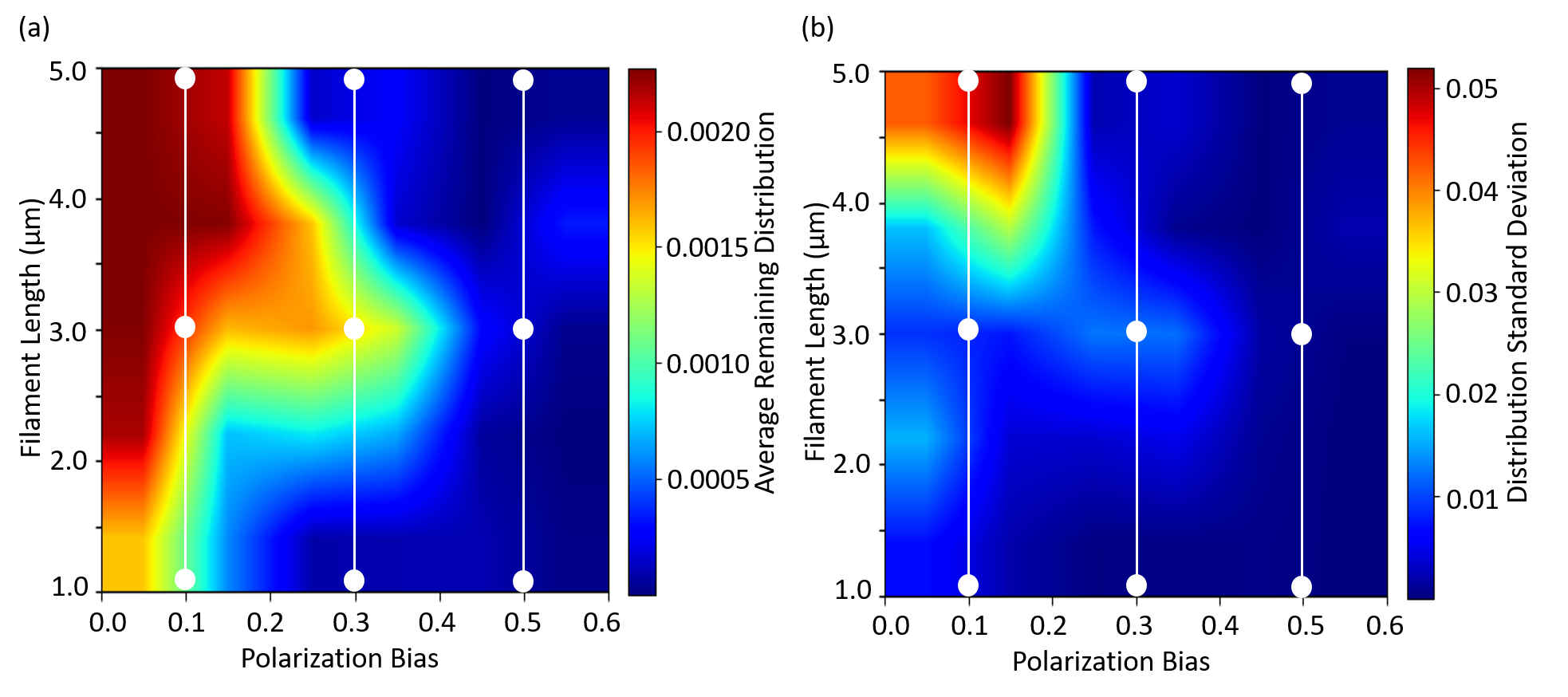}
\caption{\label{fig:survivalNinePoints} (a) The survival probability for different filament lengths and polarization biases for one network per length-bias combination. (b) The spatial standard deviation of the probability distribution after 1000 s. We will explicitly visualize the spatial distribution for nine different filament lengths and polarization biases (indicated by the white dots). It will be particularly useful to compare the distributions at the same polarization bias, but different filament lenghts (along white lines connecting the dots).}
\end{figure}

To understand the origins of the effects seen in the previous section, we hone in on the spatial distribution of the cargo probability distribution across single networks at nine different points in the filament length and polarization bias phase space, indicated by the white dots in Figs. \ref{fig:survivalNinePoints}a and \ref{fig:survivalNinePoints}b. 

Fig. \ref{fig:survivalNinePoints}a, shows the survival probability across the parameter space for a single network (as in \ref{fig:mfpts}a ). In order to quantify the extent to which traps or sequestering regions develop, we compute the spatial (point-to-point) variance in the probability distribution for the given network after 1000 s (Fig. \ref{fig:survivalNinePoints}b plots the square root of this variance). A high degree of spatial variance at late times would be indicative of a heterogeneous spatial distribution, suggesting strongly trapping regions. We see that the variance is highest when the filament length is high and the polarization bias is low. In order to understand this, we now visualize the actual spatial distribution at this point in time.

\begin{figure}[H]
\centering
\includegraphics[scale =0.35]{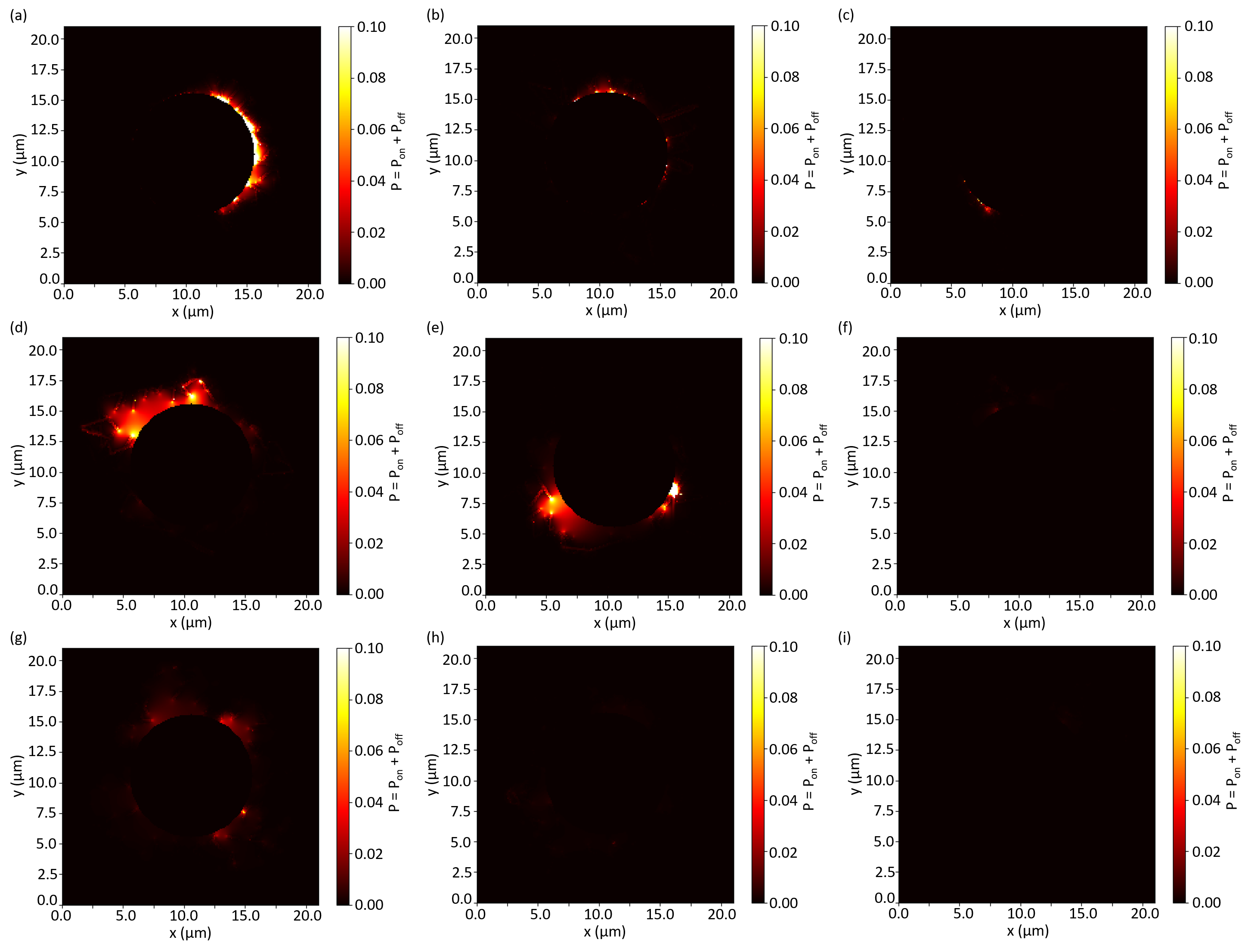}
\caption{\label{fig:nineProbDists}The cargo probability distribution after 1000 s when the filament length is 5 $\mu$m and the polariztion biases are (a) 0.1, (b) 0.3, and (c) 0.5, the filament length is 3$\mu$m and the polarization biases are (d) 0.1, (e) 0.3, and (f) 0.5, and when the filament length is 1 $\mu$m and the polarization biases are (g) 0.1, (h) 0.3, and (i) 0.5. }
%(a) shows that much of the distribution is still contained in traps near the nucleus, as this is the only area traps can occur when the filament length is 5.0 $\mu$m. Compare this with (d), where the filament length is 3.0 $\mu$m and distribution traps are more spread out within the bulk cytoplasm. This helps explain the results seen in Fig. \ref{fig:survivalNinePoints}, where at a polarization bias of 0.1, the survival probability is relatively high at filament lengths of 5.0 $\mu$m and 3.0 $\mu$m (shown in Fig. \ref{fig:survivalNinePoints}a), but the standard deviation is much lower at a filament length of 3.0 $\mu$m than it is at 5.0 $\mu$m, as can be seen by examining Fig. \ref{fig:survivalNinePoints}b. In comparing (b), (e), and (h), it can be seen that greatest amount of distribution is remaining at a filament length of 3.0 $\mu$m when the polarization bias is 0.3. This reflects the larger survival probability at intermediate filament lengths when the polarization bias is this high. In looking at (c), (f), and (i), it can be seen that most of the distribution has left the cell by this time.}
\end{figure}

Fig. \ref{fig:nineProbDists} shows the state of the probability distribution after evolving in time for 1000 s at the nine points in phase space.  In Fig. \ref{fig:nineProbDists}a, where the polarization bias is 0.1 and the filament length is 5 $\mu$m,  most of the probability distribution is still near the nucleus. This is because when the filaments are 5 $\mu$m long, they span the cytoplasm and the distribution can only be in a ``trapped" state near the nucleus. If a cargo makes it to the middle of the bulk, it will likely either be directed by filaments out of the cell, or right back to the nucleus. The latter is more likely to happen when the polarization bias is low.

The radial position of the distribution in Fig. \ref{fig:nineProbDists}a is in contrast to where the distribution appears in Fig. \ref{fig:nineProbDists}d, where, the filament lengths are 3 $\mu$m even though the polarization bias is still 0.1. Here, the distribution can be seen gathering at bright spots that are spread out even near the middle of the bulk. This is also reflected in the fact that, although the survival probability is relatively high at a polarization bias of 0.1 for filament lengths of 5 $\mu$m and 3 $\mu$m, the distribution standard deviation is much lower when the filament length is 3 $\mu$m. At these filament lengths, filament ends occur throughout the bulk and therefore trapping regions can also occur throughout the cell and correspond to locations that arise stochastically with a relatively higher number of plus ends (compared to minus ends). The filaments then serve to funnel the cargo into this region ballistically with only uninterrupted diffusion or binding to the few filaments directed away from the region allowing for escape. The occurrence of such regions was also explicitly demonstrated in a recent study of FPT across a rectangular domain with no polarization bias \cite{abel1}.  It is to be noted that these locations are regions that allow for dynamic sequestration/trapping of motile cargo without any specific binding or caging mechanism. At even smaller filament lengths of $1 \mu$m (Fig. \ref{fig:nineProbDists}g), the trapping regions are still spread out but are much less effective since imbalances between plus and minus end concentrations are less and the filaments only funnel in cargo from relatively nearby thus reducing the trapping efficiency.

For intermediate values of the polarization, the situation is even more interesting. In Fig. \ref{fig:survivalNinePoints}a, following the center white line (drawn at a polarization bias of 0.3), shows that the survival probability starts low (when the filament length is 5 $\mu$m), then {\it increases} in value (when the filament length is decreased to 3 $\mu$m), and then drops in value again (at a filament length of 1 $\mu$m). The distributions shown in Figs. \ref{fig:nineProbDists}b,e and h reflect the delayed transition of the survival probability from high to low values at intermediate filament lengths. In Fig. \ref{fig:nineProbDists}b, we can see that most of the distribution has left the cell and whatever remains is still near the nucleus, because the trapping was dependent on the long range predominantly inward flux to the nuclear boundary at lower polarizations, which is less effective at intermediate polarizations. In Fig. \ref{fig:nineProbDists}e, however, much of the distribution is still in the cell, since the trapping regions only depend on the stochastic local imbalance of plus ends that is not as strongly dependent on overall polarization. For the lowest filament lengths (Fig. \ref{fig:nineProbDists}h), the trapping is ineffective and the distribution has mostly left the cell.  Finally, for high polarizations (Fig. \ref{fig:nineProbDists}c,f,i), the outward flux dominates and the survival probability is low, independent of filament length.

\subsubsection{Conclusion and Future Directions}

In this paper, we have introduced a method to analyze intracellular transport that involves solving for the time evolution of a probability distribution of cargos moving both on and off an explicitly represented cytoskeletal network that consists of a random network of polar filaments . Numerical solutions of the associated differential equations, appropriately discretized, while losing information about individual trajectories and small number fluctuations, offer advantages over individual cargo simulations including speed and the extraction of more accurate cargo first-passage time information for large numbers of cargo. 

Using this approach, we explored the sensitivity of cargo export to the lengths of the filaments that make up the cytoskeletal network, as well as the network polarization bias. The most interesting results are seen for intermediate values of the filament lengths and polarization bias. For our choice of cell size and geometry and values of diffusion constant, ballistic speed and motor binding/unbinding kinetics, this corresponds to filament lengths that are near 3 $\mu$m and polarization bias of the network near 0.3. With this combination of network parameters, dynamic sequestering regions that maintain a relatively larger concentration of cargo for significantly longer periods of time occur throughout the bulk of the cell (Figs. 7 and 8). In this regime, the sequestering is sensitive to both filament length and polarization bias, highlighting the use of these as tuning parameters to switch between sequestration and export by increasing the polarization bias or {\it either} increasing or decreasing the filament lengths. Changing polarization bias requires overall reorientation of filaments, which would be more challenging {\it in vivo} on fast timescales, though it can be controlled by the spatial localization of filament nucleation promoters \cite{gopinathan1, heald1} on slower timescales. However, there are many cellular regulators of polymerization/depolymerization and severing of filaments \cite{gopinathan1, heald1} that could allow length to be used as a readily responsive and sensitive tuning knob for sequestration/export switching. The emergence of tunable dynamic sequestering regions in the bulk could be relevant in insulin release in pancreatic cells which transition from having restless granules that are sequestered in the bulk to exporting them upon glucose stimulation and actin depolymerization \cite{zwang, atomas, makalwat, zhu, bryan1}.  It is also interesting to note that, at lower polarization biases, filament length could also be used to switch between dispersed sequestration at intermediate filament lengths and aggregated (near the nuclear membrane) sequestration at longer filament lengths. This is consistent with the observed changes in the actin cytoskeleton between central aggregation dispersal of pigment granules in  melanophores \cite{snider1}, suggesting a role for cytoskeletal network reorganization in addition to filament switching properties of the cargo. Thus cytoskeletal filament length can serve as a regulator between qualitatively different and functionally important phases of intracellular transport.

Our approach can be extended in many useful directions. One such direction is to set up particular cytoskeletal network geometries rather than placing filaments randomly. For example, we could have microtubules that are oriented radially outward with a relatively thin layer of actin filaments near the cell membrane forming the cortex corresponding to certain geometries that have been predicted to be optimal in terms of first passage to arbitrary locations within the cell \cite{ahafner}. The method can also be easily extended to model transport on networks obtained from experimental images of real intracellular networks. Another fruitful direction is to take into account the effect of multiple-filament intersections on the transport of cargos. These intesections can cause increased molecular motor-based tug-of-war when multiple motors are present on a single cargo \cite{mjimuller, ahendricks} as well as the formation of cargo vortices and cycling behavior \cite{scholz1}. 
%These effects, along with the fact the the interior of the cell is actually dense creating crowding effects \cite{weiss1}, imply that normal diffusion is not necessarily a sufficient explanation for the passive transport phase. Incorporating anomalous diffusion into the model may, therefore, also be an interesting extension.
Proper treatment of the fate of cargo when they reach filament endpoints is also critical since this depends on the motor type and the filament polymerization rate \cite{dcai2}. Additionally, the possibilty of cargo crowding near endpoints may even help facilitate molecular motor dissociation, meaning that as cargos approach filament ends, they unbind higher rates \cite{ross2}. Finally, given that, in many cellular contexts, the cytoskeletal network can be highly dynamic (e.g. during crawling or division), incorporating the dynamics of the filaments themselves into the model will be of significant interest.

\subsubsection{Acknowledgements}
This work was supported by National Science Foundation grant NSF-DMS-1616926. AG and BM were also partially supported by the NSF-CREST: Center for Cellular and Bio-molecular Machines at UC Merced (NSF-HRD-1547848). AG would also like to acknowledge the hospitality of the Aspen Center for Physics, which is supported by National Science Foundation grant PHY-1607611, where some of this work was done. 

\bibliographystyle{unsrt}
\bibliography{numpol}

\end{document}